\documentclass[prl,twocolumn,showpacs]{revtex4}

\usepackage{graphicx}
\usepackage{amsmath}
\usepackage{amssymb}

\begin{document}
\def\a{\alpha}
\def\b{\beta}
\def\g{\gamma}
\def\e{\varepsilon}
\def\d{\delta}
\def\l{\lambda}
\def\m{\mu}
\def\t{\tau}
\def\n{\nu}
\def\o{\omega}
\def\p{\varphi}
\def\r{\rho}
\def\s{\sigma}
\def\S{\Sigma}
\def\G{\Gamma}
\def\D{\Delta}
\def\O{\Omega}

\def\ra{\rightarrow}
\def\ua{\uparrow}
\def\da{\downarrow}
\def\pd{\partial}
\def\bk{{\bf k}}
\def\bq{{\bf q}}
\def\bn{{\bf n}}
\def\br{{\bf r}}
\def\bm{{\bf m}}
\def\bp{{\bf p}}

\def\be{\begin{equation}}
\def\ee{\end{equation}}
\def\bea{\begin{eqnarray}}
\def\eea{\end{eqnarray}}
\def\nn{\nonumber}
\def\lb{\label}
\def\pref#1{(\ref{#1})}

\title{Anomalous impurity resonance in graphene}

\author{Yu.G. Pogorelov}

\affiliation{IFIMUP/Departamento de F\'{\i}sica, Universidade do Porto,
R. Campo Alegre, 687, Porto, 4169, Portugal}

\begin{abstract}
A Green function analysis has been developed for quasiparticle
spectrum and localized states of a 2D graphene sheet in presence of
different types of substitutional disorder, including vacancies. The
anomalous character of impurity effects in this system is demonstrated,
compared to those in well known doped semiconductors, and explained in
terms of conical singularities in the band spectrum of pure graphene.
The criteria for appearance of localized states on clusters of impurity
scatterers and for qualitative restructuring of band spectrum are
established and a phase diagram in variables ``disorder'' \emph{vs}
``electron density'' is proposed.
\end{abstract}
 \maketitle

There is a growing attention to electronic properties of a single
carbon layer known as graphene \cite{geim}. Its 2D honeycomb lattice
defines a peculiar band structure \cite{slon}  with two nodal points
in the Brillouin zone where conical energy surfaces (with zero effective
mass) of conduction and valence bands touch each other. This gives
rise to electronic dynamics of relativistic Dirac type \cite{sem},
extraordinary for condensed matter, and generates such unusual
phenomena as half-integer Hall effect \cite{novo,zhang,gus} and,
possibly, the magnetic catalysis of an excitonic gap \cite{mir,khve},
ferromagnetism and superconductivity \cite{kop}. On the other hand,
it is of interest to examine the effects that various kinds of
impurities can produce on this remarkable material, regarding for
instance a fundamental role of such effects in physics of common
semiconductors (with finite effective mass) \cite{shkl}. An intriguing
situation with impurity levels near conical singularities was
recognized in \emph{d}-wave superconductors \cite{lee} where
theoretical predictions are sometimes contradictory \cite{atk} and
not fully confirmed by the existing experimental data. To this time,
the disorder effects in graphene were theoretically studied, searching
for weak localization in this 2D electronic system under weak
scattering (Born limit) \cite{suz,khve1} or for strong localization
under infinitely strong (unitary limit) perturbation  \cite{paco}.
This work is aimed on a consequent description of restructured electronic
spectrum, at arbitrary perturbation strength and in a rather broad range of
impurity concentration, and on specifics of this restructuring for Dirac
quasiparticles under realistic perturbation, compared to usual quasiparticles
with parabolic dispersion.

Let us start from the simplest tight-binding Hamiltonian restricted
to nearest neighbor hopping

\be H = t\sum_{\bn, {\boldsymbol \d}}
a_{\bn}^{\dagger}b_{\bn+{\boldsymbol\d}}+h.c., \lb{eq1} \ee

\noindent where $t$ is the hopping amplitude, $a_{\bn}$ and
$b_{\bn+\boldsymbol{\d}}$ are the Fermi operators of (spinless)
electrons on sites of type 1 and 2 respectively, and atomic energy
on each site is chosen zero. The vectors $\boldsymbol \d$ point to
three nearest neighbors of a site (the vectors for a type 1 site
being the inverses of those for a type 2 site, see Fig.
\ref{fig1}a). Passing from site operators to plane waves: $a_{\bk} =
N^{-1/2} \sum_{\bn} {\rm e}^{-i \bk\cdot\bn}a_{\bn}$ and
$b_{\bk}=N^{-1/2}\sum_{\bn}{\rm e}^{-i \bk\cdot\bn}b_{\bn}$ ($N$
being the number of cells in the lattice), and then to the
eigen-modes

\begin{figure}[h!]
\includegraphics[width=8.5cm]{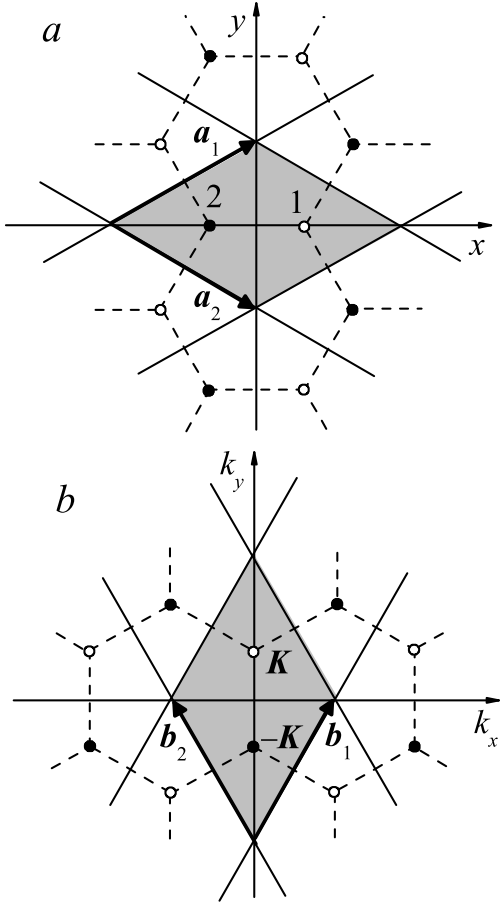}
\caption{2D lattice structure of a graphene sheet. a) Rhombic
primitive cell (shadowed) with two non-equivalent positions for
carbon atoms, 1 (open circles) and 2 (solid circles),  and
elementary translation vectors ${\bf a}_1$ and ${\bf a}_2$ of length
$a = 2.46 {\AA}$. b) Rhombic Brillouin zone (shadowed) with two
non-equivalent nodal points, ${\bf K}$ (open) and $-{\bf K}$
(solid), and vectors of reciprocal lattice ${\bf b}_1$ and ${\bf
b}_2$ of length $4\pi/(a\sqrt3)$.} \lb{fig1}
\end{figure}

\bea \a_{\bk} & = &\frac 1 {\sqrt 2} \left({\rm e}^{ - i \p_{\bk}
/2} a_{\bk} + {\rm e}^{i \p_{\bk} / 2}b_{\bk}\right),\nn\\
\b_{\bk} & = & \frac 1{\sqrt 2} \left({\rm e}^{i \p_{\bk} /2}
b_{\bk} - {\rm e}^{ - i\p_{\bk}/2}a_{\bk}\right), \lb{eq2} \eea

\noindent the Hamiltonian, Eq. \ref{eq1}, is diagonalized:

\[ H = \sum_{\bk} \e_{\bk}\left(\a_{\bk}^{\dagger}\a_{\bk} -
\b_{\bk}^{\dagger} \b_{\bk}\right),\]

\noindent with eigen-energies $\e_{\bk} = t \left|f_{\bk}\right|$.
Here the function $f_{\bk}=\sum_{\boldsymbol \d}{\rm e}^{i \bk \cdot
\boldsymbol{\d}} = \left|f_{\bk}\right|{\rm e}^{i\p_{\bk}}$ vanishes
near two isolated points in the Brillouin zone: ${\bf K} =\left(0, 2
\pi / 3 a \right)$ and $-{\bf K}$ (Fig. \ref{fig1}). Due to the
absence of inversion symmetry for the point group C$_3$, this
vanishing is linear in small difference ${\bq} = \bk - {\bf K}$ (or
${\bq} = \bk + {\bf K}$): $ f_{\bk} \equiv f_{\bq} \approx
\frac{\sqrt 3}2 \left( q_x - i q_y \right)$ defining the conical
form of isoenergetic surfaces $\e = \pm \e_{q}$ where

\be \e_{q} \approx \hbar v_{\rm F}q \lb{eq3} \ee.

\noindent with the Fermi velocity $v_{\rm F}=\sqrt 3 ta/2\hbar$. The
following analysis of this system is restricted to the low energy
physics which is essentially determined by the vicinities of two
nodal points $\pm{\bf K}$. The four relevant long-wave modes, picked
up from the eigen-modes, Eq. \ref{eq2}, near each nodal point, can
be combined into the Dirac 4-spinor $\psi_{\bq}$ whose components
(in common notation) are: $\psi_{{\bq}\ua}^{(+)} = \a_{{\bq} + {\bf
K}}$, $\psi_{{\bq}\da}^{(+)} = \a_{{\bq} -{\bf K}}$,
$\psi_{{\bq}\ua}^{(-)} = \b_{{\bq} + {\bf K}}$, $ \psi_ {{\bq}
\da}^{(-)} = \b_{{\bq} - {\bf K}}$. Here, ``particles'' and
``antiparticles'' obviously correspond to electrons and holes and
the ``Dirac spin'' indices to the nodal points (while the physical
spin indices stay suppressed). The respective Dirac form of the
Hamiltonian, Eq. \ref{eq1}, is

\be H = \sum_{\bq}\e_{q} \psi_{\bq}^{\dagger} \hat{\g}_0
\psi_{\bq}, \lb{eq4} \ee

\noindent where the 4$\times$4 matrix $\hat \g_0 = \hat
\t_3\otimes\hat\s_0$, the tensor product of Pauli matrices
$\hat\t_i$, acting on ``particle-antiparticle'' indices, and
$\hat\s_i$, on ``Dirac spin'' indices. We describe the dynamics of
this system by the (Fourier transformed) two-time Green functions
(GF's) \cite{bb}, here combined into a 4$\times$4 matrix $
\langle\langle\psi_{\bq}|\psi_{{\bq }^{\prime}}^{\dagger}
\rangle\rangle$. For the unperturbed system, Eq. \ref{eq4}, its
exact form is $\langle \langle \psi_{\bq} | \psi_{{\bq }^
{\prime}}^{ \dagger} \rangle \rangle = \hat G_{\bq}^0
\d_{{\bq},{\bq}^{\prime}}$ with

\be \hat G_{\bq}^0 = \frac{\e + \e_{q}
\hat{\g}_0}{\e^2-\e_{q}^2}. \lb{eq5} \ee

The notable distinction of graphene from genuine relativistic
systems of quantum field theory is the possibility to study the
effects of localized perturbations on its dynamics, which is our
main purpose here. To this end, we adopt the Lifshitz model of
impurity perturbation \cite{lif}, supposing a certain shift $V$ of
the on-site energy at random sites of the lattice. This model looks
more adequate to the case of rare defects in graphene, than the
alternative choice \cite{sheng} of Anderson model with random
perturbations at each lattice cite \cite{and}. We denote ${\bf p}_1$
the defect sites of type 1 with concentration $c_1$ and ${\bf p}_2$
those of type 2 with concentration $c_2$ (not necessarily equal to $c_1$),
the total impurity concentration being $c_1 + c_2 = c \ll 1$. This
perturbation scatters the modes of Eq. \ref{eq4} accordingly to the
Hamiltonian

\bea
 H^{\prime} & = & \frac V{2N} \sum_{{\bq},{\bq}^{\prime}}
\psi_{\bq}^{\dagger} \left\{ {\rm e}^{-i\p_{\bq}}
\hat U_1 {\rm e}^{i\p_{\bq^{\prime}}} \sum_{\bp_1}
{\rm e}^{i \left(\bq^{\prime} - \bq \right)\cdot \bp_1}
  \right. \nn \\
 && \left.  + {\rm e}^{i\p_{\bq}}\hat U_2 {\rm e}^{-i\p_{\bq^{\prime}}}
 \sum_{\bp_2} {\rm e}^{i\left(\bq^{\prime} - \bq \right) \cdot
\bp_2} \right\}\psi_{{\bq}^{\prime}}, \lb{eq6} \eea

\noindent with $\hat U_{1,2} =(\hat\t_0 \mp \hat \t_1) \otimes
(\hat\s_0 + \hat \s_1)$. The equations of motion for perturbed GF
matrix \cite{ilp} with Hamiltonian $H+H^{\prime}$ can be solved in
the T-matrix approximation. For the most important, momentum-diagonal
GF, this results in $\langle \langle \psi_{\bf q} |
\psi_{\bf q }^{ \dagger} \rangle \rangle = \left[\left(\hat
G_{\bq}^0\right)^{-1} - \hat \S\left(\e\right)\right]^{-1}$ where
the self-energy matrix is

\be
\hat \S\left(\e\right)= \left(c_1\hat U_1 + c_2\hat
U_2\right)\frac{ V}{2 D(\e)}.
\lb{eq7}
\ee

\noindent Its denominator $D(\e) = 1 + V g(\e)$ includes the lattice
sum $g(\e) = (2\e/N) \sum_{\bk} ( \e^2 - \e_{\bk}^2)^{-1}$ which can
be approximated at low energies, $|\e| \ll W = \sqrt{3\pi}t/2$, as:

\be
g(\e) \approx \frac \e {W^2} \ln \frac{\e^2}{\e^2 - 4 W^2/\sqrt
3}.
\lb{eq8}
\ee

\noindent A resonance can appear in $\S(\e)$ at an energy
$\e = \e_{res}$ such that ${\rm Re}\,D(\e_{res}) = 0$. But,
accordingly to Eq. \ref{eq8}, this requires a strong enough
perturbation: $|V| > V_{cr}$, where the critical value is
$V_{cr} \approx 0.947W$. For a numerical check, we adopt the
common value of $t\approx 2.5$ eV which leads to $V_{cr} \approx
3.6$ eV. The defects usually discussed in graphene are vacancies
\cite{paco}, for which the ``acceptor'' perturbation parameter
$V$ can be approximated by the 1st ionization potential of carbon
\cite{piers}: $V \approx I_C \approx 11.3$ eV $\approx 3W$. It results
in a low resonance energy: $\e_{res} \approx 0.048W \approx 0.18$ eV,
as shown in Fig. \ref{fig2}. This can be compared with the two
traditional impurities for carbon compounds: acceptors
by boron (1st ionization potential $I_B\approx 8.3$ eV) and donors
by nitrogen ($I_N \approx 14.5$ eV). The related estimates for
impurity perturbation: $V_B \approx I_C - I_B \approx 3$ eV and
$V_N \approx I_C - I_N \approx -3.2$ eV, make it less probable that
these impurities produce resonances in graphene spectrum. Therefore,
the following analysis concentrates on the effects of vacancies in
this material.

\begin{figure}
\includegraphics[width=8cm]{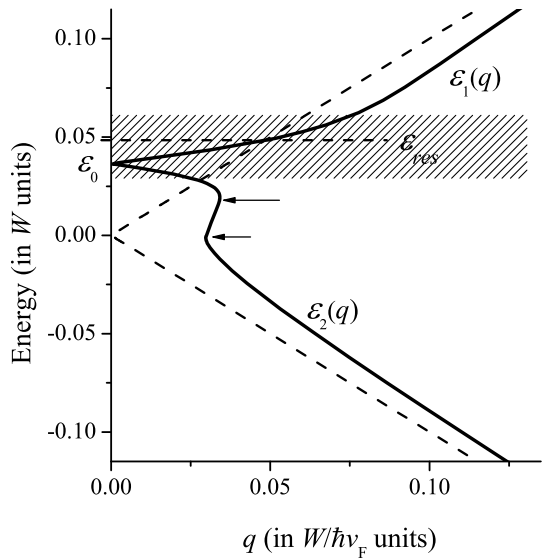}
\caption{Dispersion curves (solid lines) for graphene with acceptor
impurities at the choice of perturbation parameter $V=3W$ and
impurity concentration $c=1\%$. The dashed lines show the
unperturbed dispersion laws and impurity resonance level $\e_{res}$,
and the arrows mark the interval of negative dispersion. In fact,
the solid curves only make sense beyond the hatched area of
localized states (cf. to Fig. \ref{fig3}c).} \lb{fig2}
\end{figure}

The self-energy matrix, Eq. \ref{eq7}, can be simplified under the
most natural assumption that defects are equally present in two
sublattices: $c_1=c_2 = c/2$. Then it becomes:

\be \hat \S\left(\e\right)\to c \hat U \frac{V}{2\left[1+V
g(\e)\right]} \lb{eq9} \ee

\noindent with $\hat U = \hat\t_0\otimes\left(\hat\s_0 +
\hat\s_1\right)$, that is non-mixing particles and antiparticles.
This facilitates diagonalization of GF matrix, through a spinor
rotation: $\tilde{\psi}_{\bq} = \hat Q\psi_{\bq}$, with $\hat Q =
\hat\t_0\otimes\left(\hat\s_0-i\hat\s_2\right)/\sqrt 2$, leading it
to the form:

\bea &&\langle \langle \tilde\psi_\bq | \tilde\psi_\bq^\dagger
\rangle \rangle =\nn\\
&& \left(
                             \begin{array}{cccc}
                               \e-\e_{q} & 0 & 0 & 0 \\
                               0 & \e-\e_{q}-\S & 0 & 0 \\
                               0 & 0 & \e+\e_{q} & 0 \\
                               0 & 0 & 0 & \e+\e_{q}-\S \\
                             \end{array}
                           \right)^{-1}
\lb{eq10} \eea

\noindent with the scalar self-energy $\S(\e) = cV/[1+Vg(\e)]$. We
notice that two modes, the 1st and 3rd diagonal elements in Eq.
\ref{eq13}, stay unperturbed. These modes, $(\a_{\bq+\bf
K}-\a_{\bq-\bf K})/\sqrt 2$ and $(\b_{\bq+\bf K}-\b_{\bq-\bf
K})/\sqrt 2$, are antisymmetric in ``Dirac spin'' indices, while the
perturbation $\hat U$ is symmetric in these indices. Thus the
impurity scattering in this case perturbs only two symmetric modes,
whose dispersion is described by the equations

\be \e_{1,2}(q) - {\rm Re}\S(\e_{1,2}(q)) \pm \e_q  =  0. \lb{eq11} \ee

\noindent Using the explicit function $g(\e)$, Eq. \ref{eq8}, and
the linear law $\e_q$, Eq. \ref{eq3}, we obtain the dispersion
curves $\e_{1,2}(q)$ as shown in Fig. \ref{fig2} (the picture for ``donor''
perturbation $V<0$ will be simply inverted, $\e\to-\e$, by the
particle-antiparticle symmetry).

If Eqs. \ref{eq14} were valid down to $q\to 0$ (though in fact
they are not), they would describe the shift of nodal point energy
from zero to a finite value $\e_{1,2}(0)\equiv\e_0 = {\rm Re}\,
\S(\e_0)$. This value grows with the defect concentration as

\be \e_0 \approx \left\{\begin{array}{c}
                   cV,\qquad\qquad\qquad c\ll c_0,\\
                   \e_{res}\left(1 - c_0/c\right),\quad
                   c\gg c_0,
                 \end{array}\right.
                 \lb{eq12}
\ee

\noindent where the characteristic concentration $c_0 \sim (\e_{res}/W)^2$
defines, as will be seen below, the threshold for qualitative restructuring
of the spectrum. For $c > c_0$, there appears a certain interval of
\emph{negative} dispersion in the valence band (a Z-like feature in
Fig. \ref{fig2}). It resembles the known situation near resonances in
parabolic bands \cite{ilp}, however in this case negative dispersion appears
rather far from the resonance $\e_{res}$ (which lies within the conductance
band). And the restructured conductance and valence bands are both shifted
\emph{towards} this resonance, not repelled from it (as commonly for doped
semiconductors).

\begin{figure}
\includegraphics[width=8.5cm]{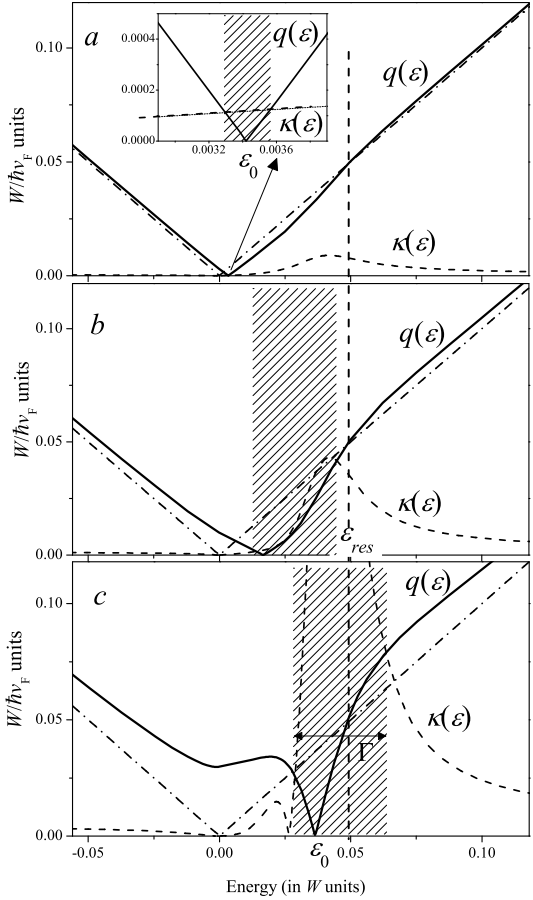}\\
\caption{Check for the IRM criterion at different concentrations of
defects. a) Below the critical concentration, $c = 0.1\% < c_0$, all
the states are band-like except a very narrow vicinity of the nodal
point energy $\e_0$ (shown under a great magnification in the
inset), where the states are localized. b) At approaching the
critical concentration, $c = 0.2\% \approx c_0$, the area of
localized states rapidly expands on the whole interval between
$\e_0$ and $\e_{res}$. c) At overcritical concentration, $c = 1\%
\gg c_0$, the area of localized states of width $\G$ extends beyond
the resonance level $\e_{res}$.}
 \lb{fig3}
\end{figure}

To clarify the physical origin of this anomalous behavior, it is
instructive to compare it with the well studied case of shallow
donor levels below a 2D parabolic conductance band $E_k = \hbar^2
k^2 /2 m$ \cite{iv}. The respective scalar self-energy formally
coincides with the above expression for $\S(\e)$ but including the
lattice sum $g(\e)=N^{-1}\sum_\bk (\e - E_k)^{-1} \approx
\ln(E_m/\e)$ and a \emph{weak} attractive perturbation
$0 < -V \ll E_m = 2\pi\hbar^2/ma^2$. Its expansion near the localized
donor level $\e_{loc} \approx -E_m {\rm e}^{-E_m/|V|}$ reads: $\S(\e)
\approx cV|\e_{loc}|/(\e - \e_{loc})$, so that the perturbed band energies
$\e = E_k + \S(\e)$ are repelled from this level. In fact, this
repulsion is due to the simultaneous action of attractive perturbation
$V$ in the numerator and denominator of $\S(\e)$ (so it remains also
true for repulsive perturbation, $V<0$, by shallow acceptors). In
contrary, such expansion for the actual case of linear band dispersion
reads: $\S(\e) \approx -cV|\e_{res}|/(\e - \e_{res})$, like if the
repulsive perturbation $V>0$ in the denominator of $\S(\e)$ turns to
be \emph{attractive} in its numerator (or vice versa for donors),
resulting in the overall attraction of the bands to the impurity
resonance.

However, the above referred dispersion curves are only reliable when
the respective band states are well defined, which can be checked by
the Ioffe-Regel-Mott (IRM) criterion \cite{irm} that the
quasiparticle lifetime is long enough compared to the oscillation
period. This can be suitably presented in the energy scale by the
inequality $q(\e) \gtrsim \kappa(\e)$ where the ``real wave number''
$q(\e)$ is the inverse function to $\e(q)$ and the ``imaginary wave
number'' $\kappa(\e) = {\rm Im}\,\S(\e)[1 - {\rm Re} \, \S^\prime
(\e)]/\hbar v_{\rm F}$. As seen from Fig. \ref{fig3}, the IRM
criterion ceases to hold just for energies close to $\e_0$, where
the quasiparticles are not properly described by the quasimomentum
and band index, but rather localized near impurity centers.
Therefore the band states in the perturbed system never include the
nodal points.

A closer insight on the localization process can be obtained, beginning
from the case of a single impurity on site $\bp$ (say, of type 1) and
constructing a quasiparticle state with energy $\e$: $|\psi_\e\rangle =
\sum_\bn \left(\psi_{\bn}^{(1)} a_\bn^\dagger + \psi_{\bn} ^{(2)}
b_\bn^\dagger \right)|0 \rangle$,  whose amplitudes on type 1 and 2 sites
$\psi_{\bn}^{(1,2)}$  are found from the Schroedinger equation,
$(H + H^\prime - \e)|\psi_\e\rangle = 0$. The solutions having central
symmetry with respect to the impurity site $\bp$:

\bea
\psi_{\bn}^{(1)} & = & \psi_\bp^{(1)} \frac {V } N \sum_\bq
\frac{{\e \rm e}^{i \bq \cdot (\bp - \bn)} } {\e^2 - \e_\bq^2} \nn\\
&=& \psi_\bp^{(1)} V\langle \langle a_\bp |a_\bn^\dagger
\rangle\rangle, \nn\\
\psi_\bn^{(2)} & = & \psi_\bp^{(1)} \frac {V } N \sum_\bq
\frac{{\e_\bq \rm e}^{i\bq\cdot(\bp - \bn)}}{\e^2 - \e_\bq^2}\nn\\
& = & \psi_\bp^{(1)} V \langle\langle a_\bp | b_\bn^\dagger \rangle \rangle.
\lb{eq13}
\eea

\noindent  are proportional to the locator Green functions. Their long
distance asymptotics (at $|\bn - \bp| \gg a$) is:

\be
\psi_{\bn}^{(1,2)} \sim \left(\frac a {|\bn - \bp|}\right)^{3/2}\cos
\left(\frac {|\bn - \bp|}{r(\e)}\right),
\lb{eq14}
\ee

\noindent with the characteristic length $r(\e) = (W a)/(\sqrt \pi
\e) \gg a$. In particular, the value $r_0\equiv r(\e_{res})$ defines
the length scale of local perturbation of quasiparticle spectrum
near a defect. So, a qualitative restructuring of this spectrum should
happen if local perturbations begin to overlap, at characteristic
concentration of defects: $c_0 \sim a^2/(\pi r_0^2) = (\e_{res}/W)^2$.
For the considered vacancy model, this value is $\sim 0.2\%$. In fact,
when the concentration $c$ reaches this level, important changes occur
in the spectrum characteristics calculated from the GF, Eq. \ref{eq10}.

\begin{figure}
\includegraphics[width=9cm]{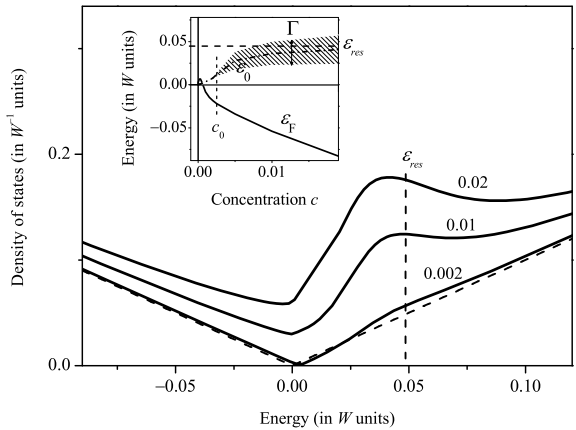}
\caption{Spectrum restructuring with growing concentration of
vacancies (indicated by numbers at solid lines for DOS $\r(\e)$).
Dashed lines show the unperturbed DOS $\r_0(\e)$ and the resonance
energy $\e_{res}$. Inset: concentration dependencies of the Fermi
energy $\e_{\rm F}$ and nodal point energy $\e_0$ (the hatched area
of width $\G$ represents its concentration broadening).} \lb{fig5}
\end{figure}

For instance, the quasiparticle density of states (DOS) is $\r(\e) =
(\pi N)^{-1} {\rm Im\,Tr}\sum_\bq \langle\langle \psi_\bq |
\psi_\bq^\dagger \rangle\rangle $. At $c \ll c_0$, it is close to
the unperturbed DOS $\r_0(\e) = (\pi N)^{-1} {\rm Im\,Tr}\sum_\bq
\hat G_\bq^0 = |\e|/W^2$, but at $c \sim c_0$ a hump appears in
$\r(\e)$ near the resonance $\e_{res}$ and then progressively
increases at $c > c_0$ (Fig. \ref{fig5}).

Next, the Fermi energy $\e_{\rm F}$ is obtained in function of $c$
from the condition $\int_{-\infty}^{\e_{\rm F}} \r(\e)d\e = (1-c)/2$,
and (in the given model) it displays a rapid initial growth at
$c \ll c_0$ (inset in Fig. \ref{fig5}), entering the conductance
band and so realizing an \emph{anomalous n-conductance at nominal
p-doping}. But, before $c$ approaches $c_0$, the level $\e_{\rm F}$
makes a sharp downturn and crosses the monotonically growing nodal
energy $\e_0$, thus restoring a usual type of conductance.

At last, the width $\G$ of the energy interval near $\e_0$ (hatched
in the inset in Fig. \ref{fig5}), where the IRM criterion ceases to
hold, is as narrow as $\sim \e_0 c/(c_0 \ln(1/c_0))\ll \e_0$ at $c
\ll c_0$ but suddenly expands at $c \sim c_0$ to the whole range
from $\e_0$ to $\e_{res}$, and then grows as $\sim \sqrt c/\ln(1/c)$
at $c \gg c_0$. This energy interval is filled by localized states,
however their localization is realized as a rule on certain clusters
of defects, rather than on single defects. On each side of this interval,
the localized states are separated from the band-like states by a Mott's
mobility edge, defining a metal-insulator transition when the Fermi
energy crosses this edge.

It is of interest to compare this type of spectrum restructuring and
the quadratic law $c_0 \sim (\e_{res}/W)^2$ with the known examples for
non-relativistic spectra. Thus, the low-frequency acoustic resonance
$\o_{res}\sim \o_{\rm D} \sqrt{M/M^\prime}\ll \o_{\rm D}$ by impurities
with mass $M^\prime\gg M$ in a crystal with atomic mass $M$ and Debye
frequency $\o_{\rm D}$ \cite{kag} gives rise to splitting of phononic
spectrum near $\o_{res}$ with opening of a quasi-gap (seen as a dip in
DOS) of width $\sim c \o_{\rm D}^2 /\o_{res}$, at surpassing the characteristic
impurity concentration $c_0 \sim (\o_{res}/\o_{\rm D})^3$ (the cubic law)
\cite{iv}. This dip  corresponds to repulsion of band levels from the
impurity resonance (in contrast to the hump in DOS and attraction of
band levels to the resonance expected in graphene). Another example is the
donor level $\e_{loc}$ near a parabolic conductance band \cite{ilps},
which rapidly expands and merges with this band when the donor concentration
exceeds $c_{loc} \sim \e_{loc} /E_m$ (linear law), in the 2D case, or $c_{loc}
\sim (\e_{loc}/E_m)^{3/2}$, in the 3D case. The 3/2 law also defines the
characteristic concentration $c_0 \sim (\o_{res}/J)^{3/2}$ of weakly
coupled ($J^\prime \ll J$) impurity spins in a Heisenberg ferromagnet
when the magnon spectrum splits near the resonance frequency
$\o_{res} \sim J^\prime \ll J$, and the cubic law $c_0 \sim (\o_{res}/J)^3$
defines such effect in an aniferromagnet \cite{ilp}. Hence the case of defects
in graphene differs from all those, even at seemingly identical linear
dispersion (as for phonons and antiferromagnons).

\begin{figure}
\center{\includegraphics[bb=300bp 10bp 440bp 580bp,
scale=0.35]{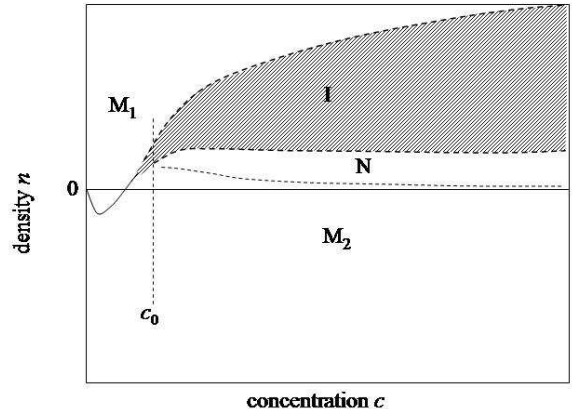}} \caption{Phase diagram of electronic states
in graphene under disorder and variable electron density (with zero level
corresponding to exact half filling.} \lb{fig6}
\end{figure}

Observable effects related to the restructured electronic spectrum of
graphene with defects need a detailed consideration which shall be
done elsewhere. Here we only outline a general framework for this
study, through the phase diagram in concentration of (vacancy)
defects $c$ \emph{vs} electron density $n$ (the latter can be
independently tuned, e.g., by the external bias voltage or by adding
non-resonant donor and acceptor impurities). Accordingly to the
concentration dependence of characteristic energy parameters in the
inset in Fig. \ref{fig5}, this diagram (Fig. \ref{fig6}) presents
three different phase regions: the ``inversely doped'' metallic
phase $M_1$ with $\e_{\rm F} > \e_0$, the insulating phase $I$ with
$\e_{\rm F}$ inside the domain $\G$ of localized states, and the
``normally doped'' metallic phase $M_2$ with $\e_{\rm F} < \e_0$.
Within the latter phase, a specific region of inverse, negative
dispersion $N$ can be distinguished, where the Fermi carriers should
behave like holes by their electric charge but like electrons by
their velocity, the latter turning extremely high near the inversion
line between $M_2$ and $N$. This peculiar dynamics can produce even
more extraordinary anomalies under applied magnetic field (which
effect is already anomalous in the pure graphene). The latter factor
and also interaction between electrons in presence of defects need a
more involved treatment.  All the mentioned features can be potentially
used for applications in a specific ``relativistic'' electronics, based on
properly doped graphene, and undoubtedly deserve further attention.

In conclusion, the effects of local perturbation by various types of
impurities, including vacancies, in single layer graphene were studied
through Green function techniques. Possibility of low energy resonance
near nodal points in the relativistic electronic spectrum was indicated
for the case of vacanicies and the conditions for qualitative restructuring
of the quasiparticle spectrum were established. A significant distinction
of this restructuring for relativistic electrons, compared to the known
disordered systems, is demonstrated and the phase diagram in variables
``disorder \emph{vs} electron density'' is proposed, indicating possible
practical applications of such system.

I would like to thank V.A. Miransky, J.P. Carbotte and S.G. Sharapov for very
useful discussions of disorder effects in graphene, and Department of
Applied Mathematics of University of Western Ontario for kind hospitality
during my sabbatical visit there, when this work was done. The support from
Portuguese Funda\c{c}\~{a}o para a Ci\^encia e a Tecnologia is gratefully
aknowledged.


\begin{thebibliography}{10}
\bibitem{geim} K.S. Novoselov, A.K. Geim, S.V. Morozov, D. Jaing,
Y. Zhang, S.V. Dubonos, I.V. Grigorieva, and A.A. Firsov,
Science \textbf{306}, 666 (2004).
\bibitem{slon}P.R. Wallace, Phys. Rev.  \textbf{77}, 622 (1947).
\bibitem{sem}G. Semenoff, Phys. Rev. Lett. \textbf{53}, 2449 (1984).
\bibitem{novo}K.S. Novoselov, A.K. Geim, S.V. Morozov, D. Jaing,
M.I. Katsnelson, I.V. Grigorieva, S.V. Dubonos,and A.A. Firsov,
Nature \textbf{438}, 197 (2005).
\bibitem{zhang}Y. Zhang, Y.-W. Tan, H.L. St\"ormer, and P. Kim,
Nature \textbf{438}, 201 (2005).
\bibitem{gus}V.P. Gusynin and S.G. Sharapov, Phys. Rev. Lett.
\textbf{95}, 146801 (2005); N.M.R. Peres, F. Guinea, and A.H.
Castro Neto, cond-mat/0506709.
\bibitem{mir}V.P. Gusynin, V.A. Miransky, I.A. Shovkovy, Phys. Rev.
Lett. \textbf{73}, 3499 (1994); E.V. Gorbar, V.P. Gusynin,
V.A. Miransky, I.A. Shovkovy, Phys. Rev. B \textbf{66}, 045108 (2002).
\bibitem{khve} D.V. Khveshchenko, Phys. Rev. Lett. \textbf{87},
206401 (2001); \textbf{87}, 246802 (2001).
\bibitem{kop}Y. Kopelevich, P. Esquinazi, J.H.S. Torres, and S. Moehlecke,
J. Low Temp. Phys. \textbf{119}, 691 (2000).
\bibitem{shkl}B.I. Shklovskii and A.L. Efros, Electronic properties of
doped semiconductors, Springer (1984).
\bibitem{lee} P.A. Lee, Phys. Rev. Lett. \textbf{71}, 1887 (1993).
\bibitem{atk}W.A. Atkinson, P.J. Hirschfeld, and A.H. McDonald,
Phys. Rev. Lett. \textbf{85}, 3922 (2001).
\bibitem{suz}H. Suzuura and T. Ando, Phys. Rev. Lett. \textbf{89},
266603 (2002).
\bibitem{khve1}D.V. Khveshchenko, cond-mat/0602398.
\bibitem{paco}N.M.R. Peres, F. Guinea, A.H. Castro Neto, cond-mat/0512091.
\bibitem{bb} V.L. Bonch-Bruevich and S.V. Tyablikov,  The Green function
method in statistical mechanics. Amsterdam: North-Holland, 1962.
\bibitem{lif} I.M. Lifshitz, Sov. Phys. JETP \textbf{17}, 1159 (1963).
\bibitem{sheng} D.N. Sheng, L. Sheng, Z.Y. Weng, cond-mat/0602190.
\bibitem{and} P.W. Anderson, Phys. Rev. \textbf{109}, 1492 (1958).
\bibitem{ilp} M.A. Ivanov, V.M. Loktev, Yu.G. Pogorelov, Phys. Rep.
\textbf{157}, 209 (1987).
\bibitem{piers} H. O. Pierson, Handbook of Carbon, Graphite, Diamond, and
Fullerenes: Properties, Processing, and Applications. New York: Noyes Data.
(1993).
\bibitem{ilps} M.A. Ivanov, V.M. Loktev, Yu.G. Pogorelov, Yu.V. Skripnik,
Sov. J. Low. Temp. Phys. \textbf{17}, 377 (1991).
\bibitem{irm} A.F. Ioffe and A.R. Regel, Progr. Semicond. \textbf{4}, 26
(1960); N.F. Mott, Adv. Phys. \textbf{16}, 149 (1967).
\bibitem{kag} Yu.M. Kagan, Ya.A. Iosilevskii, JETP \textbf{15}, 182 (1962).
\bibitem{iv} M.A. Ivanov, Sov. Phys. Solid State \textbf{12}, 1508 (1971).


\end{thebibliography}
\end{document}